# Large Linear Magnetoresistance and Shubnikov-de Hass Oscillations in Single Crystals of YPdBi Heusler Topological Insulators

Wenhong Wang[1], Yin Du[1], Guizhou Xu[1], Xiaoming Zhang[1], Enke Liu[1], Zhongyuan Liu[2], Youguo Shi[1], Jinglan Chen[1], Guangheng Wu[1], and Xixiang Zhang[3]

[1] State Key Laboratory for Magnetism, Beijing National Laboratory for Condensed Matter Physics, Institute of Physics, Chinese Academy of Sciences, Beijing 100190, China

[2] State Key Laboratory of Metastable Material Sciences and Technology, Yanshan University, Qinhuangdao 066004, P. R. China

[3] Core Labs, King Abdullah University of Science and Technology (KAUST), Thuwal 23955-6900, Saudi Arabia

We report the observation of a large linear magnetoresistance (MR) and Shubnikov-de Hass (SdH) quantum oscillations in single crystals of YPdBi Heusler topological insulators. Owning to the successfully obtained the high-quality YPdBi single crystals, large non-saturating linear MR of as high as 350% at 5K and over 120% at 300 K under a moderate magnetic field of 7 T is observed. In addition to the large, field-linear MR, the samples exhibit pronounced SdH quantum oscillations at low temperature. Analysis of the SdH data manifests that the high-mobility bulk electron carriers dominate the magnetotransport and are responsible for the observed large linear MR in YPdBi crystals. These findings imply that the Heusler-based topological insulators have superiorities for investigating the novel quantum transport properties and developing the potential applications.

Materials exhibiting large linear magnetoresistance (MR) have attracted intense research interest due to their potential applications in magnetic random access memory and magnetic sensors[1]. The MR of non-magnetic metals with open Fermi surfaces (*e.g.,* Au) can be linear and non-saturating at high fields[2]. Large linear MR effects have recently been explored in materials with approximately zero bandgaps, such as doped silver chalcogenides[3], single-crystal bismuth thin films[4], indium antimonide[5], and multi-layer graphene[6,7]. There are two prevailing models that explain the origin of linear MR behaviors. One is the quantum model by Abrikosov[8] for materials with gapless linear dispersion spectra and the other is the classical model by Parish and Littlewood (PL model)[9] for strongly inhomogeneous systems. The recent discoveries of giant linear MR in $Bi_2Se_3$[10,11] and $Bi_2Te_3$[12,13] topological insulators (TIs) has brought renewed interest to linear MR. Because of their unusual surface states that are naturally zero bandgap with linear dispersion, TIs provide a perfect platform on which to study the origin of linear MR. Because some Heusler compounds have recently been predicted to be TIs[14-16], it is worth investigating whether large linear MR can also be observed in Heusler alloys. Considerable efforts have been devoted to the study of the electric structures and transport properties of a number of Heusler-based TIs[17-25]. X-ray photoelectron spectroscopy results have shown that such Heusler-based materials as YPtSb, LaPtBi, LuPtSb and LuPdBi are gapless semiconductors with very high mobility[21,22,24] although large linear MR and desired Shubnikov-de Haas (SdH) quantum oscillations have not been observed experimentally in these materials. Among these predicted Heusler TIs is YPdBi, which stands the border between the trivial and topological insulator states. Here, we present experimental evidence demonstrating the existence of large linear MR and the first observation of SdH quantum oscillations in high-quality single crystals of YPdBi. The MR is about 350% and 120% at 5 and 300 K, respectively, under a magnetic field of 7 T. Analysis of temperature- and angle-dependence of SdH oscillations manifest that the observed large linear MR features do not come from the surface states, but arise from the high-mobility 3D bulk electron carriers in YPdBi crystals.

**Results**

**Structural characterizations.** High-quality single crystals of YPdBi were grown by the flux method as described in the Methods section. Typical dimensions of the single crystals were about 1.0 mm×1.0 mm×1.0 mm and the dimensions of the largest crystals were approximately 5 mm×4 mm×4 mm as shown in Figure 1a. Most of the crystals were located at the bottom of the crucible, which could have been caused by the high density of the materials. The quality of the single crystals was examined by the x-ray Laue back-reflection technique from different positions across one facet. The resulting Laue image obtained from a piece YPdBi single crystal, shown in Figure 1b, clearly indicates that the sample was indeed a single crystal with very well-developed (100) facets. Small triangular-like steps were found on the surfaces, which strongly suggests that the crystals grew through two-dimensional nucleation and a layer-by-layer growth mechanism on the dense planes. The composition of the crystals was charcaterized by energy-dispersive x-ray analysis (EDX) (see supplementary material Figure S1 and Table S1). The EDX results revealed that the average ratio of Y to Pd to Bi in the crystals was about 34.2 to 33.3 to 32.6, which was very close to to the stoichiometric ratio. The uncertainty of measurement was less 1.0 %, which was determined by repeating the measurements on the same sample surface. No flux component was detected. To determine the crystal structure of the samples, we ground the crystal and performed power X-ray diffraction (PXRD) measurements. Based on the PXRD patterns shown in Figure 1c, it is evident that these crytsals were single-phase crystals. The structure was found to be a MgAgAs-type cubic crystal structure (inset of Figure 1c) with lattice parameters of 6.68Å, which was consistent with previous reports on polycrystalline samples[19-21]. Moreover, the normal axis of the crystal platelets was along the (100) direction as determined by Laue diffraction. The quality of the crystal was further examined using the X-ray rocking curve, as shown in the inset of Figure 1c. The typical full-width at half-maximum (FWHM) of the X-ray rocking curve for the (200) reflection was about 0.12º, which again indicated that the crystals had good crystallinity and that the growth technique was suitable for this type of alloy.

**Temperature-dependent resistivity.** In Figure 2a and b, we show the temperature dependence of the zero-field resistivity, $\rho_{xx}$, and the carrier concentration derived from measurements of the Hall effect, assuming a one-band model. The zero-field resistivity curve shows a broad maximum at $T$~100 K and a weak temperature dependence over the whole temperature range, i.e., the d$\rho$/dT was quite small. Below the peak, the behavior of the resistivity was metallic-like, where the electron carrier density, $n_e$, was mostly temperature independent. In this constant carrier density region, the resistivity decreased with decreasing temperature owing to decreased phonon scattering. Above the peak, the resistivity decreased and the carrier density, $n_e$, increased with increasing temperature, which is commonly found in semimetal or gapless semiconductors that do not have large activation energies. The magnitude of $\rho_{xx}$ (~0.8 m$\Omega$ cm at 300 K) is consistent with the reported value of the polycrystalline YPdBi compound[18]. The high value of the carrier density ($n_e$~2.5×10$^{18}$ cm$^{-3}$ at 300 K) indicated that YPdBi is a high-carrier gapless semiconductor.

Figure 3a shows an overview of the temperature dependence of the resistivity, $\rho(T)$, of YPdBi measured under a series of applied magnetic fields. The field was applied perpendicularly to the current direction. Strikingly, the resistivity increased dramatically over a wide range of temperature, from 5 to 300 K, when a 5 T magnetic field was applied. Another interesting feature of the 5 T curve is that the overall behavior of the curve changed, e.g., the broad peak disappeared, a plateau formed below 100 K and the temperature dependence became much stronger, i.e., the dρ/dT was much larger than in the zero-field curve. In other words, the ratio of the peak value to the value at 300 K, ρ(100 K)/ρ(300 K), was about 1.3, whereas it increased to 2.1 in the 5 T curve. All these changes under a moderate magnetic field (5 T) indicate that the electron behaviors were dramatically altered by the magnetic field.

**Linear magnetoresistance.** Figure 3b shows the magnetic-field dependence of the normalized MR at different temperatures. The MR is defined as [ρ(H)- ρ(0)]/ρ(0) × 100%, where ρ($H$) and ρ(0) are the resistivities at field $H$ and zero field, respectively. A robust MR, as large as 350%, was observed under a 7 T magnetic field over a broad temperature range from 5 to 100 K. More importantly, the MR was temperature independent over this temperature range and near linearly proportional to the magnetic field. These two characteristics are essential for application of high magnetic field sensors at low temperatures in a wide temperature range. The field sensitivity was about 50%/T. Above 100 K, the MR gradually reduced with temperature, and reached to 120% at 300 K, which is also remarkable for field sensors at room temperature.

To understand the mechanism/or physics behind this giant MR, we also measured the MR when the magnetic field was parallel to the current direction, or longitudinal MR (see supplementary material Figure S2). We found that although the longitudinal MR showed non-saturating behavior, its magnitude was relatively smaller than that of the transverse MR. At temperatures below 100 K, the longitudinal MR had a weak quasi-linear dependence on the applied magnetic field and it was almost independent of temperature. At temperatures higher than 100 K, however, the quadratic contribution in the low fields became more pronounced. The longitudinal MR of the polycrystalline sample was also non-saturating, but only a very small MR of 2% was observed under a magnetic field of 7 T (see supplementary material Figure S3).

The inset of Figure 3b shows the best fitting result for the MR curve at 300 K. As a result, we found that the relation between MR and the magnetic field, $H$, could be described by the following quadratic equation: MR = $a|H| + bH^2$. From the above observations, we can conclude that the MR originated from contributions of both the linear and parabolic terms. The parabolic term is well understood as a result of the Lorentz force, whereas the origin of the linear MR remains controversial. To ascertain the origin of the observed linear MR, we first considered the quantum linear MR description proposed by Abrikosov [8] in which the zero-gap band structure with linear dispersion is important. One thus would expect that the linear MR observed in the YPdBi single crystals would be explained by the Abrikosov's quantum MR[8], as the bands of Heusler compounds exhibit linear dispersion close to the Fermi energy and the charge carriers behave like massless Dirac fermions. Therefore, the MR should be linear from low to high field and it should be independent of temperature. It is almost independent of temperature for $T$ < 100 K for magnetic fields higher than 2 T. In addition, if the linear MR was quantum mechanics in nature, then as the temperature increased, we would expect the MR to increase because the linear energy dispersion of the topological surface state can persist at both low and high temperatures. The linear MR in epitaxial grapheme [6] and topological insulator $Bi_2Te_3$ nanosheets[13] was found to increase with increasing temperature and was therefore interpreted as due to Abrikosove's linear quantum MR[7]. However, there is clear evidence that the linear MR in our crystals decreased with increasing temperature for $T > 100$ K.

**Shubnikov-de Hass oscillations.** Shubnikov-de Haas (SdH) quantum oscillations has been identified as a convincing tool for characterizing quantum transport in materials showing the three-dimensional (3D) bulk [26,27] and the two-dimensional (2D) surface states,[28-30] respectively. We therefore carried out low-temperature and high-field magnetoransport measurement to provide experimental evident for the quantum transport in YPdBi crystal. The magnetic field is perpendicular to both the current flow and the surface of the YPdBi crystal. The magnetic-field dependent resistivity $R_\perp$ shows traces of SdH oscillations in our raw data (See supplementary Figure S4), reflecting the high carrier mobility. In Figure 4a, we show the oscillatory component of $\Delta R_\perp$ versus $1/H$ at various temperatures ($T$) after subtracting their linear background (See supplementary Figure S4). The amplitude of the SdH oscillations decreases with increasing $T$, and the oscillations are not observed for $T > 20$K. Moreover, a single oscillation frequency can be extracted from fast Fourier transform (FFT) spectra ($f_{SdH}(T)=45$ T, See supplementary Figure S5). The frequency directly gives a Fermi surface cross-section area of $S_F=0.037$Å$^{-2}$. The tilt angle between $H$ and the surface of crystal can be varied from 90° to 0°, and we find the SdH oscillations can be also observable at 0° (See supplementary Figure S6), suggesting the oscillations are essentially due to the Landau quantization of the 3D bulk Fermi surface. Assuming a spherical Fermi surface, the carrier concentration at 2 K was estimated to be $1.7\times10^{18}$ cm$^{-3}$, which is close to that estimated by the Hall analyses, so we can ascribe the origin of the SdH oscillation not to the possible surface states but to the intrinsic gapless bulk nature of YPdBi. In addition, the temperature dependence of the SdH oscillation amplitude can be fitted to the standard Lifshitz-Kosevich theory,[31]

$$\Delta R(T,H) \propto \exp[-2\pi^2 \kappa_B T_D/\Delta E_N(B)] \frac{2\pi^2 \kappa_B T/\Delta E_N(B)}{\sinh[2\pi^2 \kappa_B T/\Delta E_N(B)]} \quad (1)$$

In Eq. (1), $\Delta E_N$ and $T_D$ are the fitting parameters, and $H$ is the magnetic field position of the Nth minimum in R. $\Delta E_N(B) = heH/2\pi m^*$ is the energy gap between the $N$th and $(N+1)$th LL, where $m^*$ is the effective mass of the carriers, $n$ is the electron charge, and $h$ is the Planck constant. $T_D = h/4\pi^2 \tau \kappa_B$ is the Dingle temperature, where $\tau$ is the quantum lifetime of the carriers due to scattering, and $\kappa_B$ is Boltzmann's constant. In Figure 4b, we show the temperature dependence of SdH oscillation amplitude at 9T. The solid line shows the best fit to Eq. (1). The inset of Fig. 4b shows the calculated $\Delta E_N$ as a function of $H$. The slope of a linear fitting yields a rather small cyclotron mass $m^* \sim 0.09\, m_e$ ($m_e = 9.1 \times 10^{-31}$ kg is the mass of free electron). The Dingle temperature $T_D$ is found to be 6K from the slope in the semilog plot of $\Delta R/R(0) H \sinh(2\pi^2 \kappa_B T/\Delta E_N)$ vs. $1/H$ at T=2K [Figure 4c]. From $T_D$ we extract the carrier scattering time $\tau = 4 \times 10^{-13}$ s. This value of $\tau$ corresponds to the carrier mobility of 7800 cm$^2$V$^{-1}$s$^{-1}$ at 2K, in reasonable agreement with the value of the Hall mobility, which provides a strong, quantitative argument that the SdH oscillations arise from the 3D bulk states.

**Discussion**

To understand the origin of the observed large linear MR, we recalled the classical PL model [9], in which the linear MR is expected to be governed by carrier mobility. The classical PL model predicts that the crossover field, $H$c (the field at which the MR curve becomes linear), is in inverse proportion to the carrier mobility, $\mu_e$, and that it should continually increase with increasing temperature due to decreasing mobility. In Figure 5a, we show the $H$c and carrier mobility, $\mu_e$, as a function of temperature. The $H$c was obtained from the first-order derivative of the MR curve with field as plotted in the inset of Fig. 3a. The carrier mobility, $\mu_e(T)=R_H(T)/\rho(T)$, was extracted from the Hall coefficient, $R_H(T)$, using a single band model. Here, we see that $H_c$ remains somewhat constant for $T < 100$ K, whereas it continually increases with increasing temperatures for $T > 100$ K. Moreover, we found that the Hall mobility, $\mu_e$, decreased slightly with temperature, although the magnitude was not strongly dependent on the temperature. The PL model [8] also predicts that the slope of the linear part of the MR, i.e., $dMR(T)/dH$, is proportional to the carrier mobility, $\mu_e$. Figure 5b displays a plot of $dMR(T)/dH$ as a function of $\mu_e$. Interestingly, we found that $dMR(T)/dH$ depends crucially on carrier mobility. The higher the mobility, the larger the $dMR(T)/dH$ magnitude. We suggest that MR and $\mu_e$ qualitatively follow the relation $MR(T) = \mu_e(T)H$, which has been reported in some recent works on polycrystalline samples[20-22]. We can therefore conclude that the linear part of MR of Heusler compounds is mainly controlled by mobility, particularly in high temperature regions. According to the classical PL MR model[9], linear MR is a consequence of mobility fluctuations. Local probe techniques showed that electron-hole puddles [26] and charge imhomogeneity [27] were found in high-quality graphene in which the Dirac electronic dispersion spectrum is firmly established. We therefore cannot rule out the possibility that strong electronic inhomogeneity is responsible for the linear MR in our YPdBi single crystals.

Let us summarize the important features of high-quality YPdBi crystal. First, it shows a large non-saturating MR as high as 350% at 5 K and over 120% at 300 K under a magnetic field of 7 T. The large linear MR exhibited an interesting dependence on the temperature, and the magnitude of the MR increases markedly as mobility increases at T < 100K. Second, it has a gapless semiconductor band structure with high-mobility 3D bulk carriers. The high-mobility and relatively low carrier concentration are the key factors for the observation of SdH oscillations at low temperature. Third, the high-mobility electron conduction is rather sensitive to the randomly oriented domains, as evidenced by the absence of SdH oscillations in the polycrystalline Heusler TIs with ultrahigh mobility carriers [22]. To reveal the origin of the high electron mobility, we have performed band-structure calculations for YPdBi as shown in supplementary Figure S7. The results are consistent with previous calculations [14], and at the equilibrium lattice constant (a=6.68Å) a linear and gapless (Dirac-cone-like) energy dispersion is formed by the $\Gamma_8$ hole-like and the highly dispersive electron-like $\Gamma_6$ bands. Moreover, as pointed out by Chadov *et al.*[14], a very small variation of the lattice constant can further shift down the parabolic dispersion heavy-hole $\Gamma_8$ band, leaving the single Driac cone at the Fermi energy, which allows all conduction carriers to condensate at the lowest Landau level. These features are compatible with the observation of the SdH oscillations, suggesting an isotropic Fermi surface in YPdBi. Nevertheless, a precise direct observation of Fermi surface by angle-resolved photoemission spectroscopy (ARPES) is desired to indentify the band structures at Fermi surface in YPdBi.

## Methods

**Sample preparation.** Single crystals of YPdBi were grown in a two-step process. The first step was to prepare polycrystalline samples of YPdBi by arc melting stoichiometric amounts of the constituent elements in a high-purity argon atmosphere. Y (99.95%) pieces, Pd (99.99%) grains, and Bi (99.999%) grains were used as starting materials. An excess of 3% Bi was used to compensate for the loss of elements during the arc melting. For better homogeneity and crystallinity, the arc-melted ingots were wrapped in Tantalum foil and then annealed in evacuated quartz tubes at 1073 K for two weeks. The second step was to grow crystals of YPdBi by a flux method. The fabricated polycrystalline YPdBi was grounded, mixed with fluxed Pb pieces in an atomic ratio of 1:10, and placed in a Tantalum crucible, which was loaded into a fused quartz tube. For estimation of the best ratio of the fluxed material to YPdBi, we examined the mixtures with various weight ratios, ranging from 20:1 to 1:1. The 10:1 mixture resulted in the best solubility and crystal growth. The tube was sealed under Ar gas at the pressure of $10^{-4}$ Torr and then placed in a furnace. We also tried to estimate the best cooling rate. The best growth conditions at the 10:1 ratio were as follows: the mixed sample was heated at 900 ºC and then at 1300 ºC for 15 hours each and then slowly cooled to 1030 ºC at a rate of 3 ºC/h.

**Materials characterization.** The composition of the single crystal samples was determined by energy-dispersive X-ray (EDX) spectroscopy and the structure were checked by X-ray diffraction (XRD) with Cu-*K*α radiation. The single-crystal orientation was checked by a standard Laue diffraction technique.

**Transport measurements.** We milled the single-crystal samples to a thickness of 0.20±0.01 mm for the transport measurements. Electrical leads were attached to the samples using room-temperature cured silver paste by gold wires. The in-plane resistivity, $\rho_{xx}$, and Hall resistivity of a selected crystal were all measured using the four-terminal method with a dc-gauge current of 5000 μA between 2 and 300 K, using a commercial Physical Properties Measurement System (PPMS) from Quantum Design. The perpendicular and longitudinal magnetoresistances were measured with the magnetic field perpendicular or parallel to the electrical current direction. Magnetoresistance versus magnetic field was also measured at several temperatures. The Hall effect was measured by rotating the crystal by 180° in a magnetic field of 5 T in PPMS between 2 and 300 K. The Hall coefficient was calculated from the slope of the measured Hall effect curves.

**Band structure calculations.** The electronic structure calculations in this work were performed using full-potential linearized augmented plane-wave method, as implemented in the package WIEN2K.[34] The exchange correlation of electrons was treated within the local spin density approximation (LSDA) including Spin-orbital coupling (SOC). Meanwhile, a 17×17×17 $k$-point grid was used in the calculations, equivalent to 5000 k points in the first Brillouin zone. Moreover, the muffin-tin radii of the atoms are 2.5 a.u, which are generated by the system automatically. The lattice parameters and atomic positions were taken from our experimental data.

## Acknowledgements

This work was supported by funding from the "973" Project (2012CB619405) and NSFC (Nos. 51071172, 51171207 and 51025103). We are grateful to Prof. X. C. Ma and Dr. R. Shan for fruitful discussions. Dr. Virgina Unkefer improved the English.

## Author contributions

W.H.W. conceived the idea and supervised the overall research. W. H. W and Z. Y. Liu designed the experiments. W.H.W. and Y.G.S grew the single crystals. J.L.C. carried out the Laue diffraction measurement. W.H.W., Y.D, G.Z.X. carried out the resistivity and Hall measurements. X.M.Z. and E.K.L. performed band calculations. W.H.W., E.K.L. and G.H.W. analyzed the data and wrote the manuscript with input from all other co-authors.

## Additional information

The authors declare no competing financial interests.

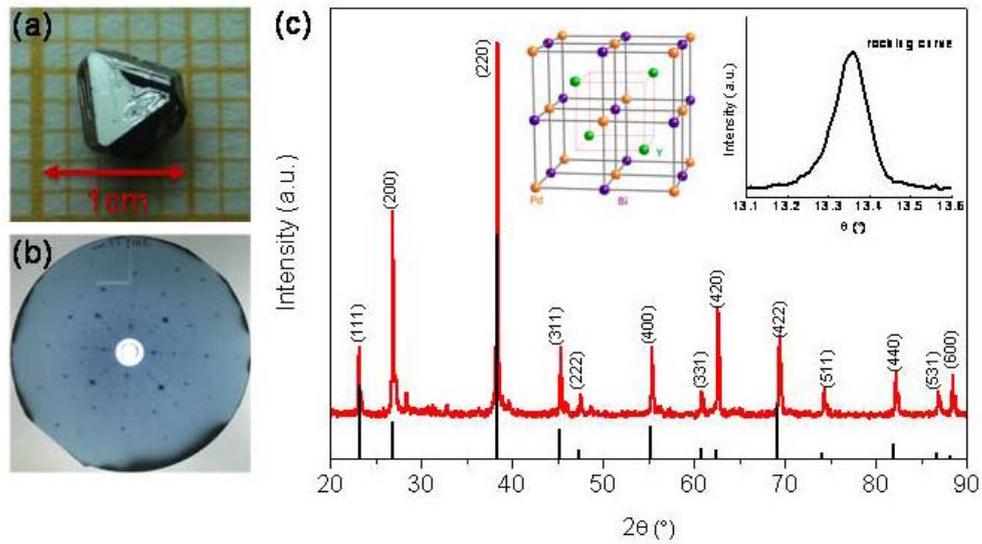

Figure 1| (a) Photograph of a single crystal of YPdBi. (b) Laue diffraction pattern of sing-crystal YPdBi generated with the beam axis coniccident with the [100] zone axis at room temperature. (c) Powder XRD pattern of pulverized single-crystal YPdBi sample, together with the simulated XRD data for fully ordered YPdBi with MgAgAs-type half-Heusler structure. Left inset shows a structure view of half-Heusler and right inset shows the X-ray rocking curve of the as-grown YPdBi single crystals, (200)- Bragg reflection.

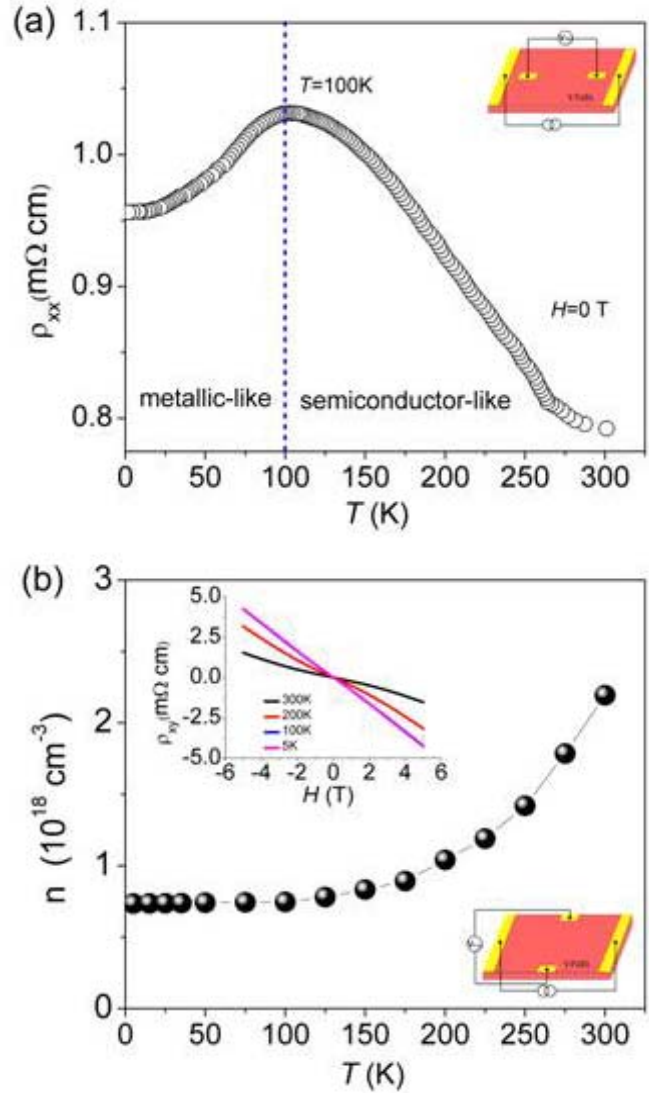

Figure 2| (a) Temperature-dependent zero-field resistivity $\rho_{xx}$. In the region of constant carrier concentration, 5K < $T$ < 100K, the resistivity decrease with decreasing temperature owing to decreased phonon scattering. (b) Variation of carrier concentration with temperature deduced from Hall effect measurement. The inset shows the Hall resistivity $\rho_{xy}$ of YPdBi as a function of magnetic field. The dominant charge carriers are electrons.

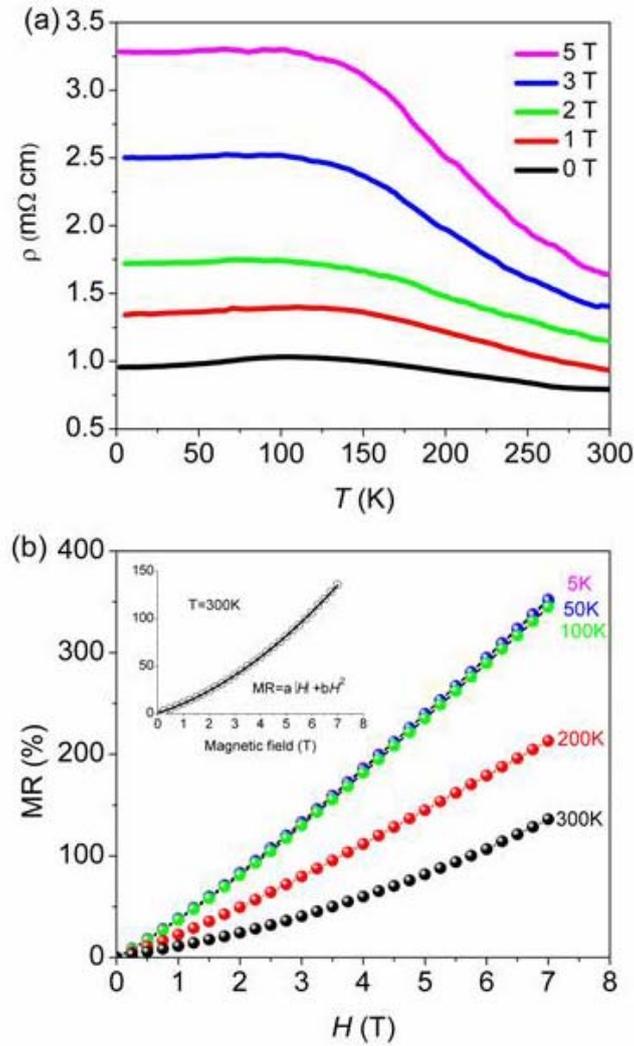

Figure 3| (a) Temperature-dependent resistivity of the single-crystal of YPdBi at a series of magnetic fields $H$=0, 1, 2, 3, 4 and 5 T applied fields. (b) Normalized MR as a function of magnetic field H at a series of temperatures T=5, 50, 100, 200 and 300 K. Here MR=[ρ($H$)-ρ(0)]/ρ(0)]× 100%, where ρ($H$) and ρ(0) are the resistivity with and without the magnetic field H, respectively. Inset, the observed MR of YPdBi at 300 K (open circles) and fit (black line) with the quadratic equation MR = $a|H| + bH^2$.

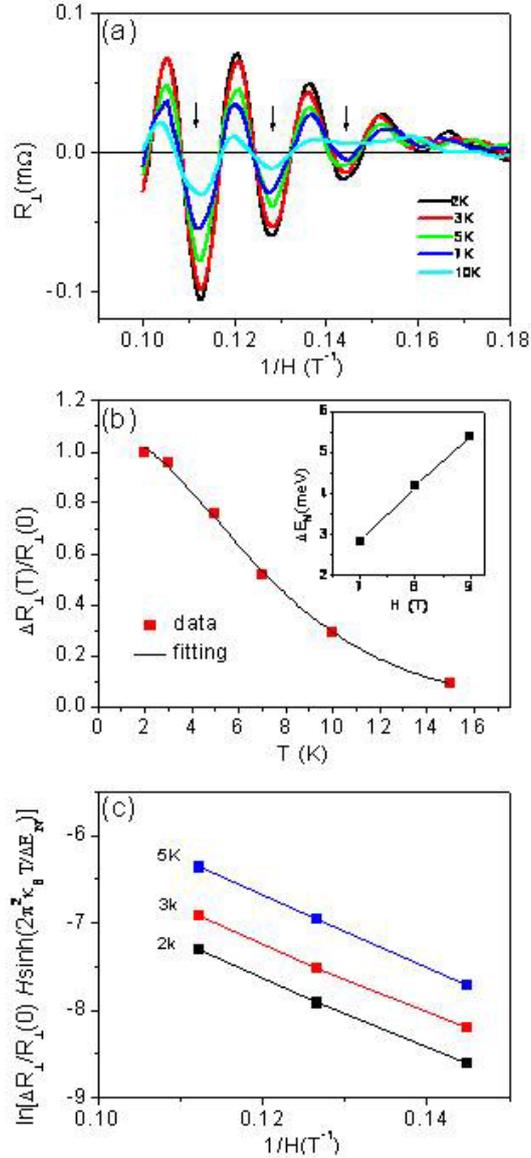

Figure 4| SdH oscillations in YPdBi crystal. (a) The variations of $\Delta R_\perp$ vs. $1/H$, after subtracting a fitted smooth background at various temperatures. (b) Temperature dependence of the SdH oscillation amplitude $\Delta R_\perp (T)/R_\perp$ at 9T. The solid line is a fit to the Lifshitz-Kosevich formula, from which we extract Landau level (LL) energy gap $\Delta E_N$. The inset shows $\Delta E_N$ as a function of $H$ (magnetic field positions of minima in $R_\perp$ corresponding to different LLs), and the effective mass:~0.09 $m_e$ is extracted from the slope of the linear fitting. (c) Dingle plots of $\ln[\Delta R/R(0) H \sinh(2\pi^2 \kappa_B T/\Delta E_N)]$ versus $1/H$ at different temperatures. The Dingle temperature $T_D$=6 K is calculated from the slope of the linear fit, corresponding to a carrier lifetime $\tau \sim 4 \times 10^{-13}$ s and an effective mobility of 7800 $cm^2 V^{-1} s^{-1}$.

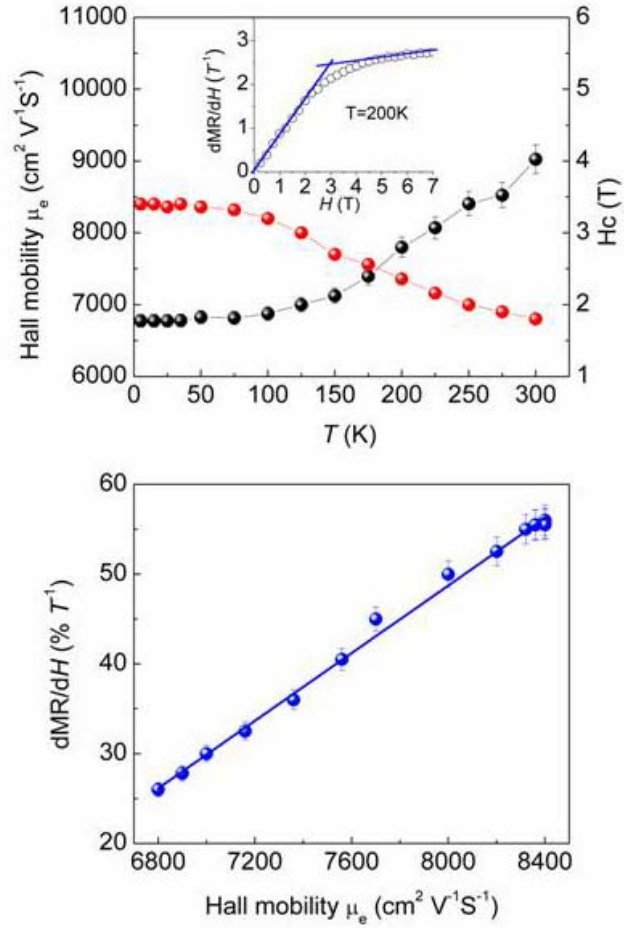

Figure 5| (a) The crossover field $H_C$ on left axis and Hall mobility $\mu_e$ on right axis as a function of temperature. Error bars are standard deviation. Inset shows the first derivative of MR of YPdBi with field at $T$=200 K, the intersection of linear indicates a turning point between parabolic and linear MR. (b) Linear part of the slope of the MR, i.e., $d$MR($T$)/$dH$ as a function of carrier mobility $\mu_e$. Error bars are standard deviation.

# Supplementary Materials to

Large Linear Magnetoresistance and Shubnikov-de Hass Oscillations in Single Crystals of YPdBi Heusler Topological Insulators

W. H. Wang [1]*, Y. Du[1], G. Z. Xu[1], X. M. Zhang[1], E. K. Liu[1], Z. Y. Liu [2], Y. G. Shi[1], J. L. Chen[1], G. H. Wu[1], and X. X. Zhang[3]

[1] State Key Laboratory for Magnetism, Beijing National Laboratory for Condensed Matter Physics, Institute of Physics, Chinese Academy of Sciences, Beijing 100190, China

[2] State Key Laboratory of Metastable Material Sciences and Technology, Yanshan University Technology, Qinhuangdao 066004, P. R. China

[3] Core Labs, King Abdullah University of Science and Technology (KAUST), Thuwal 23955-6900, Saudi Arabia

*To whom correspondence should be addressed. E-mail: wenhong.wang@iphy.ac.cn

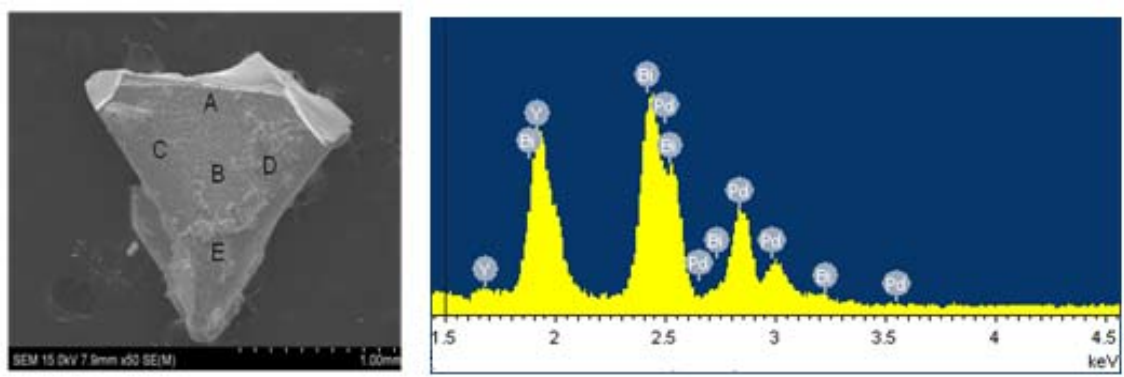

**Figure S1| A scanning electron microscope with energy-dispersive X-ray spectrometry (SEM-EDX) with a thin beryllium window system was used to determine the composition of YPdBi single crystals.** As shown in **Table S1**, the composition is fairly uniform in different regions. No flux component was detected by SEM-EDX.

**Table S1. Elemental composition of a single crystal of YPdBi obtained by EDX at different positions (see Figure S1).**

| positions | Y (at.%) | Pd (at.%) | Bi (at.%) |
|---|---|---|---|
| A | 34.53 | 33.23 | 32.24 |
| B | 34.00 | 33.52 | 32.46 |
| C | 34.08 | 33.59 | 32.33 |
| D | 33.78 | 33.27 | 32.95 |
| E | 34.45 | 32.83 | 32.72 |

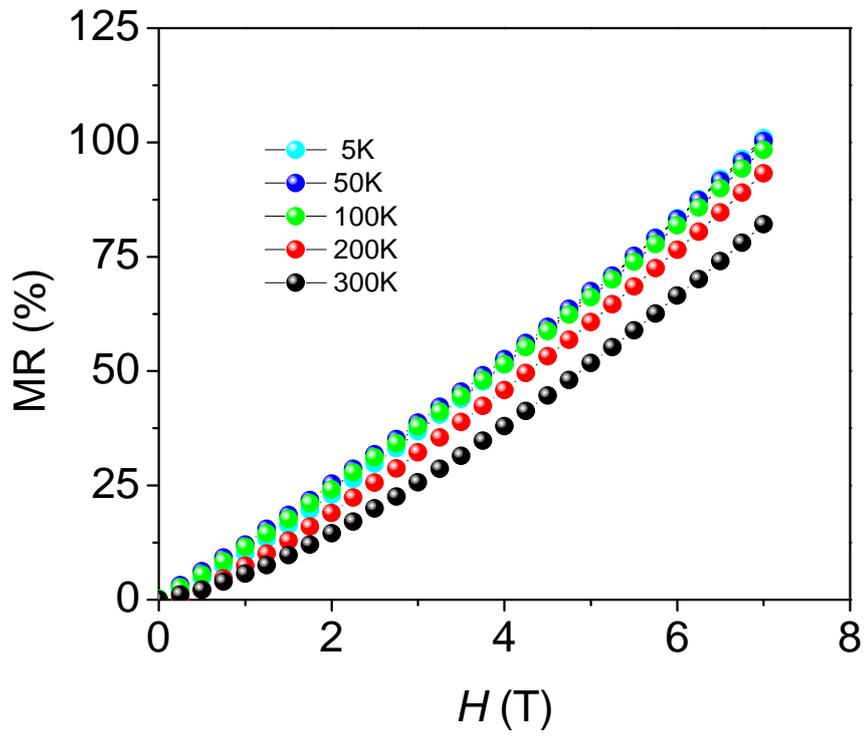

**Figure S2|** **The normalied MR for the single-crystal YPdBi samples with the magnetic field parallel to the current direction measured at different temperatures.**

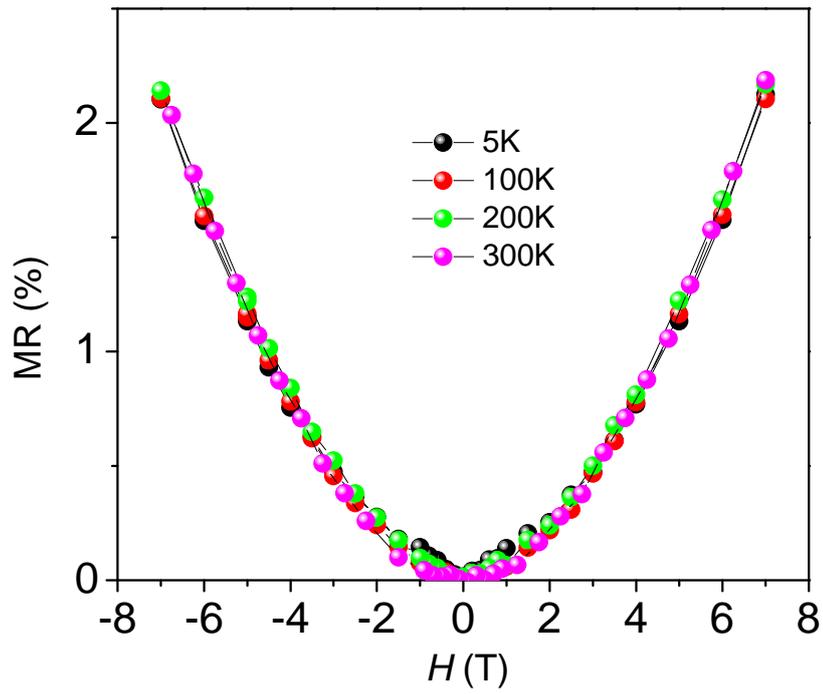

**Figure S3| The normalied MR for polycrystalline YPdBi samples with the magnetic field perpendicular to the current direction.** A nearly temperature-independent MR of 2% was observed under a magnetic field of 7 T.

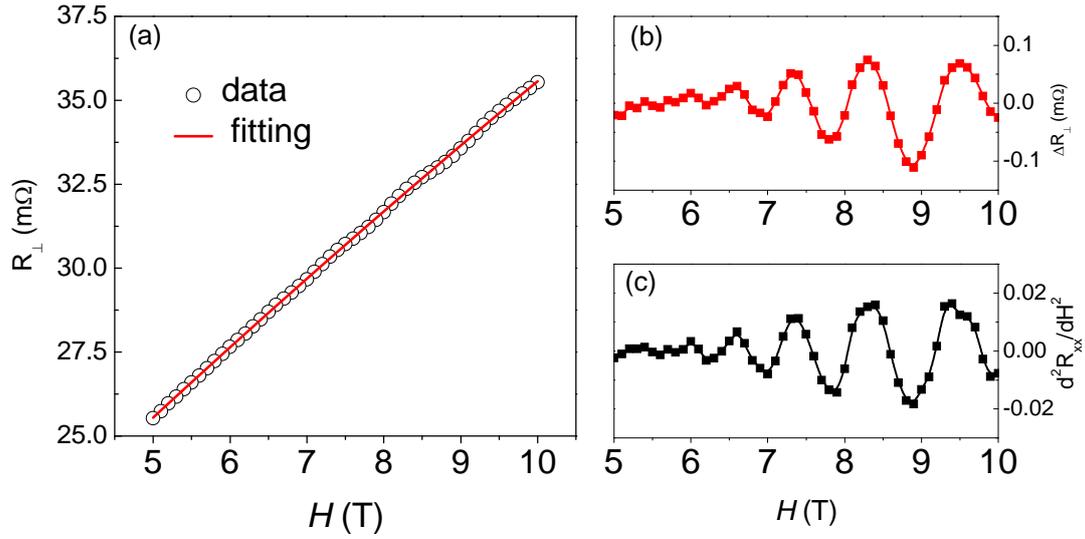

**Figure S4| The extraction of quantum oscillations from the YPdBi transport measurements.** The magnetic field ($H$) is perpendicular to both the current flow and the surface of the YPdBi crystal. (a) Magnetic field dependent MR under $H$ = 5~10 T and $T$ = 2 K. A linear background can be subtracted to obtain the oscillatory part of MR, as displayed in (b) and Figure 5a in the manuscript. Here, in FigureS4 (c), a plot of $d^2R_\perp/dH^2$ as a function of H, revealing the evident SdH oscillations and confirming the valleys (minima) are located at the same positions. This plot provides an alternative approach to separate the oscillatory part of the MR from the background. It is consistent with the results from direct subtraction method as shown in (b).

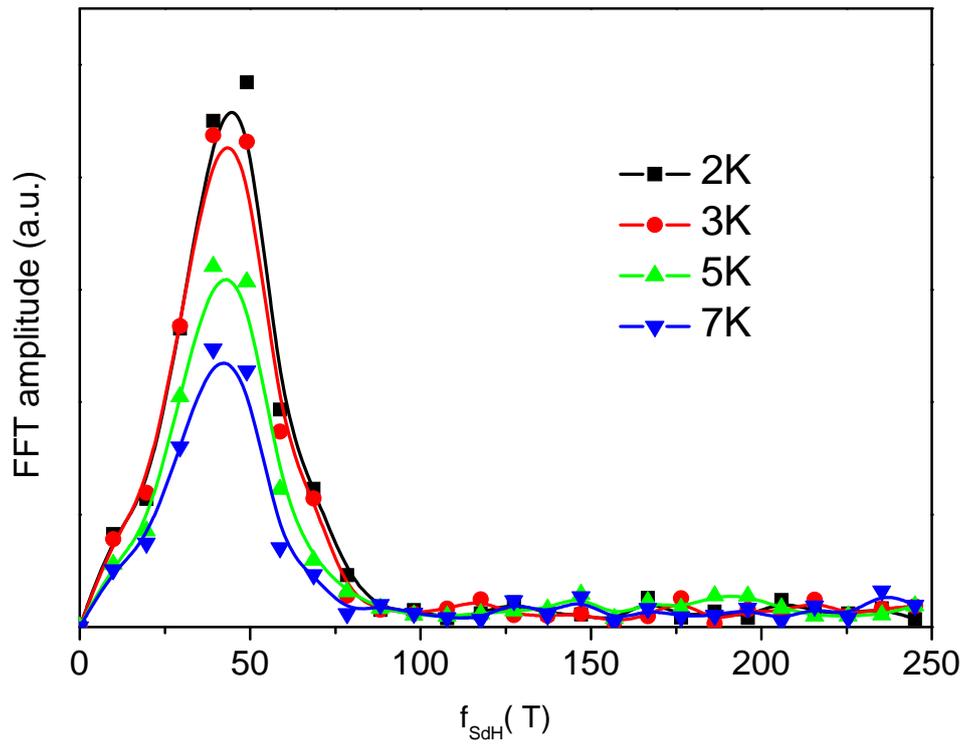

**Figure S5| Fast Fouier transform (FFT) spectra of SdH oscillations.** A single oscillation frequency, $f_{SdH}(T)$ ~45 T, can be inferred from the spectra.

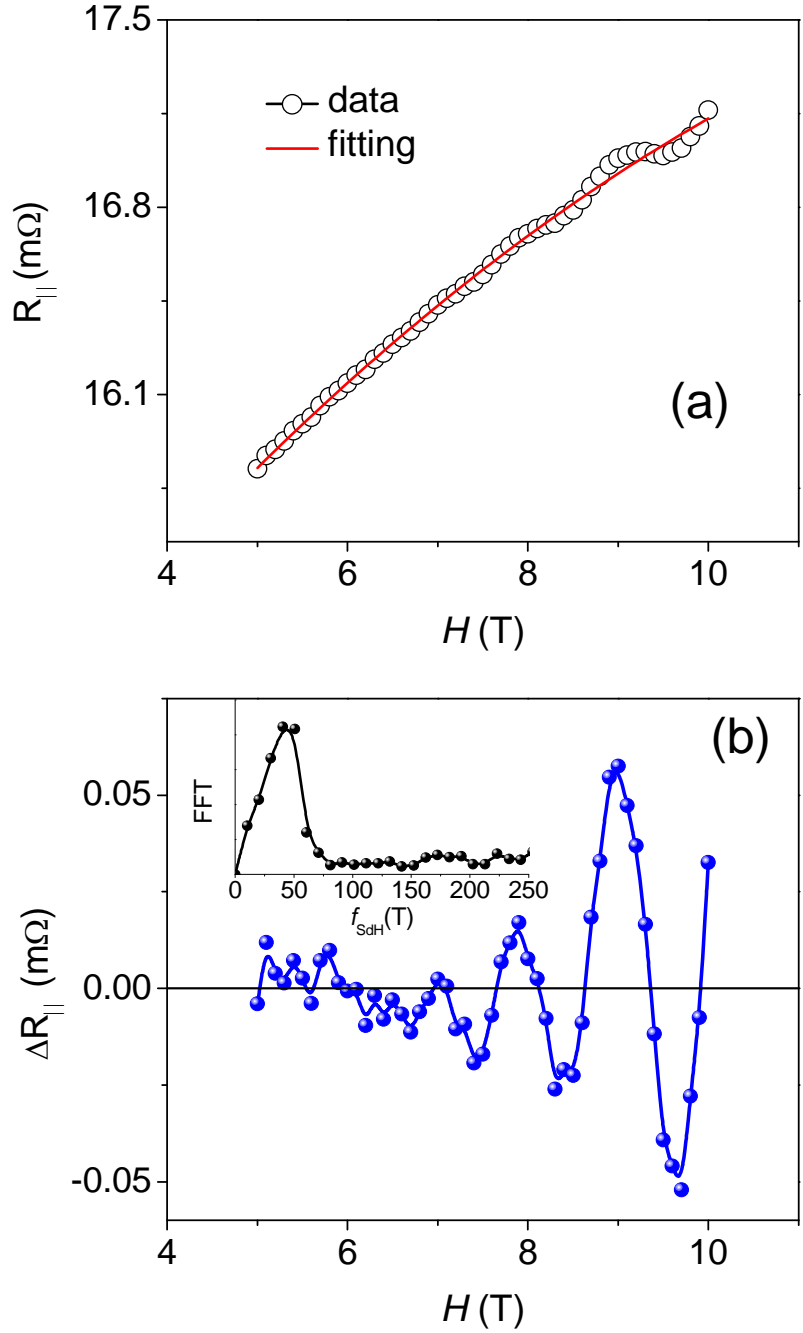

**Figure S6| The extraction of quantum oscillations from the YPdBi transport measurements.**

The magnetic field ($H$) is perpendicular to the current flow but parallel to the surface of the YPdBi crystal. (a) Magnetic field dependent MR under $H$ = 5~10 T and $T$ = 2 K. A linear background can be subtracted to obtain the oscillatory part of MR, as displayed in (b). The inset of (b) shows the FFT spectrum of SdH oscillation, and a single oscillation frequency, $f_{SdH}$(T) ~44.9 T, can be obtained.

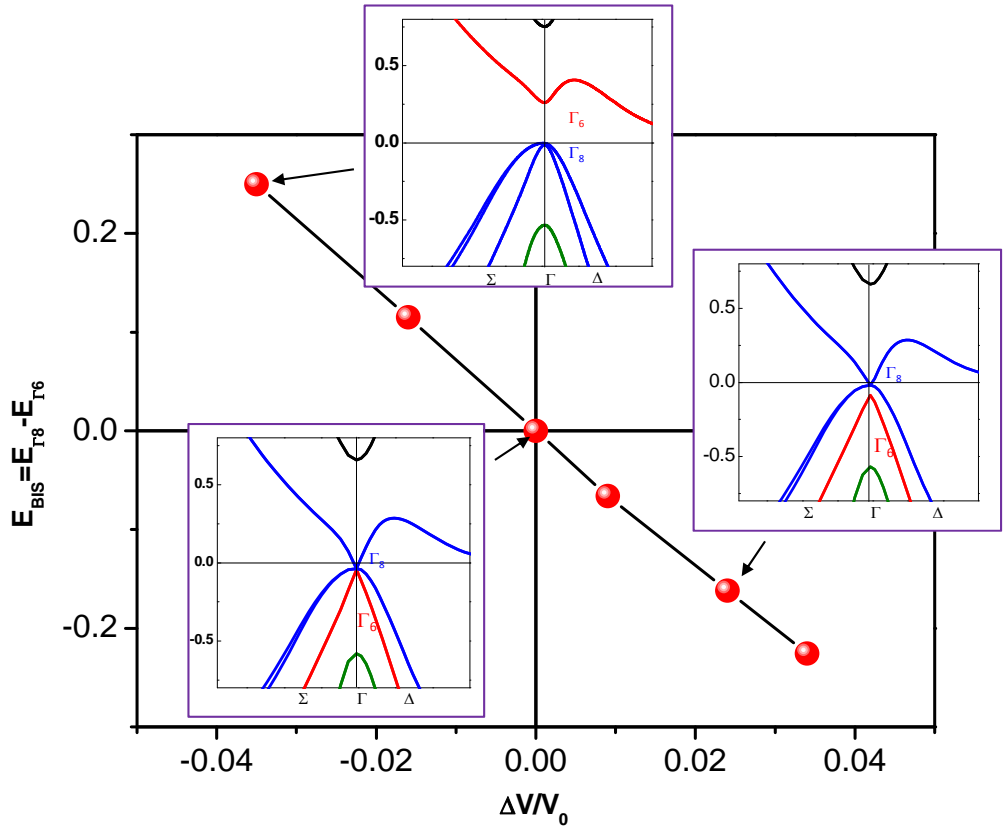

**Figure S7|** $E_{BIS}$ (=$\Gamma_6$-$\Gamma_8$) as a function of the volume changes. The insets show the band structures at the marked points along $\Sigma$ and $\Delta$ symmetry directions of the Brillouin zone. These results are well agreement with the previous calculations (Ref. 14 in the manuscript). At the experimental lattice constant ($\Delta V/V_0$=0), marked by a dashed vertical line, a linear and gapless (Dirac-cone-like) energy dispersion is formed by the $\Gamma_8$ hole-like and the highly dispersive electron-like $\Gamma_6$ bands.